\documentstyle[psfig,epsfig,onecolumn]{mn}
\def\mnras{MNRAS}
\def\apj{ApJ}
\def\apjl{ApJ}
\def\etal{{\it et al.\ }}
\def\grad{{\bf \nabla}} 
\def\fo{f(\Omega)}
\def\xib{\bar{\xi}}
\def\fosq{f^2(\Omega)}
\def\ss{(s_{\perp},\st)}
\def\s{{\bf s}}
\def\r{{\bf r}}
\def\ro{{\bf r_1}}
\def\rw{{\bf r_2}}
\def\sp{{\bf s}^{'}}
\def\ssp{(s^{'}_{\perp},\spt)}
\def\sps{(s_{\perp},\spt)}
\def\spp{(s_{\perp},\st+\st^{'})}
\def\st{s_{\parallel}}
\def\spt{s^{'}_{\parallel}}
\def\n{{\bf n}}
\def\v{{\bf v}}
\def\k{{\bf k}}
\def\kt{k_{\parallel}}

\def\vp{V_P}
\def\sig{\sigma_P^2}
\def\sigm{\sigma_P}
\begin{document}
\title[Nonlinear redshift distortions]{Nonlinear redshift
distortions: The two-point correlation function}
\author[Somnath Bharadwaj]{Somnath Bharadwaj \\ Department of Physics
and Meteorology and  
 Centre for Theoretical Studies \\ Indian Institute of Technology,  
Kharagpur,   721 302, India \\ email: somnath@cts.iitkgp.ernet.in} 
\date{}
\maketitle
\begin{abstract}
We consider a situation where the density and peculiar velocities in
real space are linear, and we calculate $\xi_s$ the two-point
correlation function in redshift space, incorporating all non-linear
effects which arise as a consequence of the map from real to redshift
space. Our result is non-perturbative and it includes the effects of
possible multi-streaming in redshift space. We find that the
deviations from the predictions of the linear redshift distortion
analysis increase for the  higher spherical harmonics of
$\xi_s$. While the deviations are insignificant for the monopole
$\xi_0$, the   hexadecapole $\xi_4$ exhibits large deviations from the
linear predictions. For a COBE normalised $\Gamma=0.25$, $h=0.5$ CDM
power spectrum  our results for $\xi_4$  deviate from the linear
predictions by a factor of two at the scales   $\sim 10 h^{-1} {\rm
Mpc}$. The deviations from the linear predictions depend separately on
$\fo$ and $b$. This holds the possibility of removing the degeneracy
that exists between these two parameters  in the linear analysis of
redshift surveys which yields only $\beta=\fo/b$. 

We also show that the commonly used phenomenological model where the 
non-linear redshift two-point correlation function  is
calculated by convolving the linear redshift correlation function
with an isotropic pair  velocity distribution function is  
a limiting case of our result. 
\end{abstract}
\section{Introduction}
It has been long  recognized that galaxy peculiar motions introduce
distortions in the  clustering pattern observed in redshift surveys.   
On small scales, the random motions of galaxies in virialized clusters
causes structures to appear elongated along the line of sight, often
referred to as the Fingers-of-God. On large scales, the coherent
inflow of galaxies into over dense structures causes these to appear
flattened along the line of sight (Peebles 1980,\S 76, and references
therein).   

In a seminal paper Kaiser (1987) showed that in the linear regime the
galaxy peculiar velocities are expected to introduce a quadrupole
anisotropy in  $P_s({\bf k})$  the power-spectrum in galaxy redshift
surveys, the relation between  $P_s({\bf k})$ and its real space
counterpart  $P_r(k)$ being
\begin{equation}
P_s({\bf k})=(1+ \Omega_0^{0.6} \mu^2)^2 P_r(k) 
\label{eq:in1}
\end{equation}
where $\mu$ is the cosine of the angle between ${\bf k}$ and the line
of sight.  
He  proposed that this effect could be used to
measure the cosmological density parameter $\Omega$. This possibility
has led to a lot of work on Linear Redshift Distortions and the reader
is referred  to Hamilton (1998) for a comprehensive review. Although
in principle Kaiser's original proposal is quite simple, there are a
large number of problems which arise in the actual analysis of
redshift distortions.

One of the main problems arises from the possibility  that galaxies
may be biased tracers of the underlying mass distribution which
determines the peculiar velocities. The fact that galaxies of
different types cluster differently 
(e.g., Dressler 1980; Lahav, Nemiroff \& Piran 1990, 
Santiago \& Strauss 1992; Loveday \etal 1995;
Hermit \etal 1996; Guzzo \etal 1997) implies that not all of them can
be exact indicators of the mass distribution. The presence of bias is
a consequence of the complex process of galaxy formation and this
suggests a  scale dependent,  stochastic and nonlinear bias
(e.g. Cen \& Ostriker 1992, Mann, Peacock \& Heavens 1998,  Ue-Li Pen
1998, Dekel \& Lahav 1999). Various studies (e.g. Taruya \etal 2001,
Benson \etal  2000, Kauffmann, Nusser \&  Steinmetz 1997) indicate
that deterministic linear biasing is a reasonably good assumption  
at large scales where the density fluctuations are linear. In this
situation biasing is described using  only one scale independent 
parameter $b$ which relates the fluctuation in the galaxy
density and the mass density through a linear
relation $\delta_g=b \delta_m$. It then   follows that the
analysis of  linear redshift distortions allows the determination of
the quantity $\beta=\Omega^{0.6}/b$, and not $\Omega$ and $b$
individually. The analysis of different 
redshift surveys has given values of $\beta$ which vary
considerably in the range $0.2 \, - \, 1.1$ (Table. 1 of Hamilton
1998). A part of the large spread in values is related to
uncertainties in the nature and extent of bias in the different
galaxy samples. The parameter $\beta$ can also be measured
by comparing the observed peculiar velocities with the observed
galaxy distribution (eg. Strauss and Willick, 1995).  The spread in
the  values of $\beta$ determined using this method  (Table 3. of Strauss
and Willick, 1995) is comparable to that 
in the values determined using redshift distortion.
Another method of measuring $\beta$ is based on the observed cluster
abundance (Wu, 2000).  Melott \etal (1998) have proposed that the
``Bulls-eye effect'' which arises due to redshift distortion offers a
different way of probing $\Omega$.   
 
 Kaiser's original work and much of the subsequent work
is based on  the plane-parallel-approximation (PPA) which is
valid if the angles  subtended by  the pairs of galaxies 
in the analysis are small. The radial nature of the redshift
distortions has to be taken into account when analyzing wide angle
surveys. Different strategies for incorporating this effect have been
studied by    Fisher, Scharf \& Lahav (1994), 
Heavens \& Taylor (1995),  Hamilton \& Culhane (1996),  Zaroubi \&
Hoffman (1996), Szalay, Matsubara \& Landy (1998)and   Bharadwaj (1999). 
The  radial gradients of the galaxy selection function
and the motion of the observer introduce new effects when the radial
nature of the redshift distortions are taken into account. These two
effects do not contribute in the plane parallel approximation. 
 
The linear analysis of redshift distortions, applicable on large scales 
where the real space density is linear  $(\sigma_l \le 1)$,
incorporates 
the distortion due to peculiar velocities only to  linear order when
transforming from real to redshift space.  
N-body simulations suggest that non-linear effect are significant
in redshift space even at considerably large scales  where the  real 
space density field is linear 
(e.g. Suto \& Suginohara 1991; Fisher et   
al. 1993; Graman, Cen \& Bahcall, 1993; Brainerd et al., 1996;
Bromley, Warren \& Zurek, 1997). This is also seen in the two-point
correlation function measured in redshift surveys. In addition to the
flattening predicted by the linear analysis, the two-point correlation
function remains elongated  along the   line of sight   
at scales as large as $20 h^{-1} {\rm Mpc}$ (e.g. Figure 2. Peacock
\etal 2001). 
This indicates that there is a regime where the density and peculiar
velocity fields in real space are  adequately described by linear
theory, and it is the mapping from real to redshift space which is
non-linear. This is often  referred to as the translinear regime.
It is important to incorporate the non-linear effects in the
analysis of redshift distortions in this regime and  
there have been two different approaches to this problem.
	
The first approach (e.g. Fisher \etal 1994, Peacock \& Dodds 1994,
Ballinger, Peacock \& Heavens 1996)  ) is phenomenological and the effect
of the non-linearity is incorporated by 
convolving the linear redshift two-point correlation function 
with the line of sight component of a random isotropic pairwise
velocity distribution function (eq. \ref{eq:dis1}). Equivalently, these
effects are introduced into the   power spectrum through a function
$\hat{f}(k,\mu)$ which multiplies the linear redshift power
spectrum.  The
resulting redshift power spectrum is given by 
\begin{equation}
P_s({\bf k})=(1+ \beta \mu^2)^2 P_r(k) \hat{f}(k,\mu)\,.
\end{equation}
Different forms, including a Gaussian and an exponential,
have been used for the distribution function   and this approach is
found to match the results of N-body simulations well (Hatton \&
Cole  1998). 

The other approach is based on the Zel'dovich approximation. 
Taylor \& Hamilton (1996), Fisher \& Nusser (1996) and Hatton and Cole
(1998) have used the Zel'dovich approximation to analytically study
the behaviour of the redshift-space power spectrum in the translinear
regime. They find that the results from Zel'dovich approximation are
in agreement with N-body simulations in predicting the shape of the
quadrupole to monopole ratio for the redshift power spectrum. They
also demonstrate that there can be departures from the linear
predictions even at relatively large scales. The first two groups of
authors  also conclude that the agreement
between the Zel'dovich approximation and N-body simulations suggests
that it is the coherent infall into over dense structures and not 
random motions in virialized clusters which is responsible for 
departures from the linear behaviour in the translinear regime.
 Recently Hui, Kofman \& Shandarin (2000) have used the
Zel'dovich approximation to study the probability distribution
function of density in redshift space. They find that the
peculiar velocities may significantly increase the multi-streaming
in redshift space even when this is insignificant in real space. This 
suggests  that non-perturbative effects may be important in the
mapping from real to redshift space even when the real space density
is relatively linear. 


It is the aim of this paper to analytically study the redshift
two-point  correlation function in the translinear regime focusing on
the  
non-linear effects  which arise from the  mapping from real
to redshift space. It is hoped that such a study will help 
elucidate the relation between the redshift correlation function and
the different real space quantities on which it depends. 
We  consider a situation where the density fluctuations and peculiar
velocities are 
linear in real space, and we calculate $\xi_s$  the two-point
correlation function in redshift space taking into 
account all non-linear effects of the redshift distortion.  
The calculation closely follows the
method used to calculate the real space two-point correlation function
in the Zel'dovich approximation (Bharadwaj 1996b). Our result is 
non-perturbative and hence it  incorporates the effects of
multi-streaming  in redshift space. Our calculation assumes the 
plane-parallel approximation and hence the effect of gradients in 
the selection function and the motion of the observer have been
ignored.  The calculation is presented in section \S 2. 
 
In \S 3 we compare our results with the predictions  of linear
redshift distortion and investigate the  nature and extent of the
deviations from the linear results. 
The effects of linear  redshift distortion are quantified by a single
parameter $\beta=f(\Omega)/b$, and it not possible to determine 
$\Omega$ and $b$ separately. We investigate whether  this degeneracy
is broken and if it becomes possible to determine $\Omega$ and
$b$ if the non-linear redshift distortions are taken into account. 

In \S 4 we present the  summary and  discussion. 

\section{Calculating $\xi_s$}
The position of a galaxy in redshift space $\s_1$ differs from its 
actual position $\r_1$,  the relation between the two being
\begin{equation}
\s_1 - U_1 \, \n_1 = \r_1
\label{eq:a0}
\end{equation}
where $U_1=\v_1 \cdot \n_1$ is the line-of-sight component of the
peculiar 
velocity of the  galaxy and  units have been chosen 
so that the 
Hubble  parameter $H=1$. We study how the  galaxy two-point
correlation function in phase space $\rho_{2r}(\r_1,\r_2,\v_1,\v_2)$ 
is related to its redshift space counterpart
$\rho_{2s}(\s_1,\s_2,\v_1,\v_2)$.

 The mapping from real space to
redshift space preserves the number of galaxies   
\begin{equation} 
\rho_{2s}(\s_1,\s_2,\v_1,\v_2) d^3 s_1  d^3 s_2  d^3 v_1  d^3 v_2 
=\rho_{2r}(\r_1,\r_2,\v_1,\v_2) d^3 r_1  d^3 r_2  d^3 v_1  d^3 v_2   
\label{eq:a1}
\end{equation}
For  pairs of galaxies which are at a large distance from the observer and
subtends a small angle in the sky, the 
 Jacobian of the
transformation from $\s$ to $\r$ (eq. \ref{eq:a0}) may be neglected
and we get
\begin{equation} 
\rho_{2s}(\s_1,\s_2,\v_1,\v_2) 
=\rho_{2r}(\s_1-U_1 \n,\s_2-U_2 \n,\v_1,\v_2)
\label{eq:a2}
\end{equation}
in the ``plane parallel approximation''. Here the unit vector $\n$ refers
to the common line-of-sight to the pair of galaxies, and the
line-of-sight components of the peculiar velocity of the two galaxies are
taken to be parallel.

The phase space two-point  correlation function in real space is 
homogeneous which allows us to write equation (\ref{eq:a2})
as 
\begin{equation}
\rho_{2s}(\s_1,\s_2,\v_1,\v_2)=\rho_{2r}(\s-U \n,\v_1,\v_2)
\label{eq:a3}
\end{equation}
where $\s=\s_2-\s_1$ is the vector joining the pair of galaxies and  
$U=\n \cdot (\v_2-\v_1)$ is the line-of-sight component of the difference 
in peculiar velocities of the two galaxies.  The vector $\s$ is
decomposed into two parts 
\begin{equation}
\s=\s_{\perp} + \st \,  \n \,,
\end{equation} 
$\st$ the component along the line-of-sight and $\s_{\perp}$ the part
perpendicular to the line-of-sight.

A Taylor expansion of equation (\ref{eq:a3}) in powers of $U$ gives us 
\begin{equation}
\rho_{2s}(\s_1,\s_2,\v_1,\v_2)=\sum_{m=0}^{\infty} \frac{(-U)^m}{m!} 
\frac{\partial^m}{\partial \st^m} \rho_{2r}(\s,\v_1,\v_2) \,.
\label{eq:a4}
\end{equation}
 We  calculate $\xi_s\ss$ the galaxy two-point correlation function in
redshift space by integrating out  the 
velocity information in equation (\ref{eq:a4}) 
\begin{eqnarray}
\bar{n}_g^2 [1+\xi_s\ss]&=&  \int d^3 v_1 d^3 v_2
\rho_{2s}(\s,\v_1,\v_2)  \nonumber \\
&=&  \int d^3 v_1 d^3 v_2
\sum_{m=0}^{\infty} \frac{(-U)^m}{m!} 
\frac{\partial^m}{\partial \st^m} \rho_{2r}(\s,\v_1,\v_2) \,.
\label{eq:a5}
\end{eqnarray}
where $n_g$ is the mean number density of galaxies and  $s_{\perp}=\mid
\s_{\perp} \mid$. Equation (\ref{eq:a5})
relates the galaxy two-point correlation in redshift space to a sum of  
velocity moments of different orders in real space. The velocity
moments in real space can also be expressed as 
\begin{equation}
\int d^3 v_1 d^3 v_2 U^m  \rho_{2r}(\s,\v_1,\v_2) = \bar{n}_g^2 
\langle U^m (1+\delta_{1r}) (1+\delta_{2r}) \rangle
\label{eq:a6}
\end{equation}
where $\langle \rangle$ denotes ensemble average, and  $\delta_1$ and
$\delta_2$ refer to the perturbation in the galaxy number density at
the position $\s_1$ and $\s_2$ in real space. Using this and 
separating the odd and  even powers of $U$ in equation (\ref{eq:a5})  
 we obtain
\begin{eqnarray}
1+\xi_s\ss&=&  \sum_{q=0}^{\infty} \frac{1}{(2q)!} 
\left(\frac{\partial}{\partial \st }\right)^{2q} 
\langle U^{2q} (1+\delta_{1r}) (1+\delta_{2r}) \rangle \nonumber \\
&-& \sum_{q=0}^{\infty} \frac{1}{(2q+1)!} 
\left( \frac{\partial}{\partial \st} \right)^{2q+1} 
\langle U^{2q+1} (1+\delta_{1r}) (1+\delta_{2r}) \rangle \,.
\label{eq:an1}
\end{eqnarray}

Our analysis is restricted to a situation where  the peculiar
velocities  and density perturbations are linear in
real space. It is also assumed that these quantities are a
Gaussian random field. We next discuss some of the statistical  
properties of $U$ and $\delta$ in real space. 

The ensemble average of an odd number of $U's$ and $\delta's$ is
zero. For example
\begin{equation}
\langle U \rangle=\langle \delta_{1r} \rangle =\langle U \delta_{1r}
\delta_{2r} \rangle  =0 \,.
\label{eq:a7}
\end{equation}

The quantities in terms of which we  express all the velocity moments
encountered in equation (\ref{eq:a5}) are
the real space galaxy two-point correlation
\begin{equation}
\xi_r\ss=\langle  \delta_{1}\delta_{2} \rangle \, ,
\label{eq:a8}
\end{equation}
the line-of-sight component of the pair velocity 
\begin{equation}
\vp\ss=
\langle U (\delta_{1}+\delta_{2}) \rangle  =\n \cdot
\langle (\v_{2}-\v_{1}) 
(\delta_{1}+\delta_{2}) \rangle \, ,
\label{eq:a9}
\end{equation}
and the dispersion of the line-of-sight component
of the pair  velocity 
\begin{equation}
\sig\ss=
\langle U^2  \rangle  =\n_i \n_j 
\langle (\v_{2}-\v_{1})_i   (\v_{2}-\v_{1})_j  \rangle \,.
\label{eq:a10}
\end{equation}

We also have the relation 
\begin{equation}
\vp\ss=
2 \,\langle U \delta_{1r} \rangle 
= 2 \, \langle U \delta_{2r} \rangle \,.
\label{eq:a11}
\end{equation}
which follows from the statistical isotropy and homogeneity of the
peculiar velocities and perturbations.  

All the higher velocity moments can be expressed in terms of
$\xi_s$, $\vp$ and $\sig$.  In addition, these three quantities are
not independent and they can be related using linear perturbation 
theory (\S 4).    

We first consider the term $\langle U^{2q} \rangle$ which can be
expressed as 
\begin{equation}
\langle U^{2q} \rangle = \frac{(2q)!}{2^q q!} [\sig\ss]^q \,.
\end{equation}

We next consider the even  velocity
moments  in equation (\ref{eq:an1}). These can be expressed as (for $q >0$)
\begin{eqnarray}
& &\langle U^{2q} (1+\delta_{1r}) (1+\delta_{2r}) \rangle 
= \label{eq:a12} \\
&=& \langle U^{2q} \rangle \,  \langle (1+\delta_{1r}) (1+\delta_{2r})
\rangle 
+  (2q)(2q-1) \langle U^{2q-2}\rangle \, \langle  U \delta_{1r} \rangle 
\, \langle U \delta_{2r} \rangle  \nonumber \\
&=& \frac{(2q)!}{2^q q!} [\sig\ss]^q [1+\xi_r\ss]   
+  \frac{(2q)!}{2^{q-1} (q-1)!} [\sig\ss]^{q-1} \frac{\vp^2\ss}{4} \,, 
\nonumber
\end{eqnarray}
and for the odd moments we have 
\begin{eqnarray}
\langle U^{2q+1} (1+\delta_{1r}) (1+\delta_{2r}) \rangle &=&
(2q+1) \langle U^{2q}  \rangle \,  \langle U (\delta_{1r}+\delta_{2r})
\rangle  
 \nonumber \\
&=&\frac{(2q+1)!}{2^{q} q!} [\sig\ss]^{q} \vp\ss \,.
\label{eq:a13}
\end{eqnarray}

Using these for the odd and even terms  in equation (\ref{eq:an1})
 we get 
\begin{eqnarray}
1+\xi_s\ss&=&\sum_{q=0}^{\infty}
\frac{1}{2^q \, q!} \frac{\partial^{2q}}{\partial \st^{2q}} \left(
 [\sig\ss]^q [1+\xi_r\ss]     \right. \label{eq:a14} \\  
&-& \frac{\partial}{\partial \st} \{ [\sig\ss]^q \vp\ss \}
+ \left.  \frac{\partial^2}{\partial \st^2} \{ [\sig\ss]^q
\frac{\vp^2\ss}{4} \}  \right)   
\nonumber
\end{eqnarray}

Each of the three terms in the above equation is next expressed using 
a Dirac Delta function.
\begin{equation}
\delta^3(\s-\sp)= \int \frac{d^3 k}{(2 \pi)^3} 
\exp[i (\s-\sp)\cdot \k ] \,.
\end{equation}

Below we show this procedure for the third term in equation
(\ref{eq:a14}). 
\begin{eqnarray}
& &\frac{\partial^{2q+2} }{\partial \st^{2q+2}}
\{ [\sig(\s)]^q \frac{\vp^2(\sp)}{4} \}  =
\frac{\partial^{2q+2} }{\partial \st^{2q+2}}
\int  d^3 s^{'} \delta^3(\s-\sp)
\{ [\sig(\sp)]^q \frac{\vp^2(\sp)}{4} \} \nonumber \\
&=& \int  d^3 s^{'} \frac{\vp^2(\sp)}{4}
\frac{\partial^{2} }{\partial \st^{2}}  \int \frac{d^3 k}{(2 \pi)^3} 
\exp[i (\s-\sp)\cdot \k ] [-\kt^2 \, \sig(\sp)]^q 
\label{eq:a15}
\end{eqnarray}
where $\kt=\k \cdot \n$ is the line of sight component of $\k$. 

Carrying out a similar procedure for all the terms in  (\ref{eq:a14})
we obtain 
\begin{eqnarray}
 1+\xi_s\ss
&=&\int d^3 s^{'} \left[ \xi_{r}\ssp + \left(1  - \frac{\vp\ssp}{2}
\frac{\partial}{\partial \st} \right)^2  \right] \times \nonumber \\
&\times&
 \int \frac{d^3 k}{(2 \pi)^3} \exp[i (\s-\sp)\cdot \k ] 
 \sum_{q=0}^{\infty}
\frac{1}{2^q \, q!}  [- \kt^2 \, \sig\ssp]^q  \,.
\label{eq:a16}
\end{eqnarray}

The sum over $q$ gives a Gaussian in $\kt$ which can be integrated
\begin{equation}
\int \frac{d^3 k}{(2 \pi)^3} \exp[i (\s-\sp)\cdot \k ] 
\exp\left[ -\frac{\kt^2
\sig\ssp}{2} \right] 
 = \delta^2(\s_{\perp}-\sp_{\perp})  G(\st-\spt,\sigm\ssp)
\,,
\label{eq:a18}
\end{equation}
where we  use 
\begin{equation}
G(x,a)=\frac{1}{\sqrt{2 \pi} a} \exp[- \frac{x^2}{2 a^2}]
\end{equation}
to represent a normalised Gaussian distribution. 

 Using this in equation (\ref{eq:a16}) we can integrate over 
 $d^2 s_{\perp}$  to get
\begin{equation}
1+\xi_{s}\ss =  \int   d \st^{'} \,
\left[\xi_{r}\sps + \left(1  - \frac{\vp\sps}{2}
\frac{\partial}{\partial \st} \right)^2  \right]
G(\st-\spt,\sigm\sps)
\end{equation}

Replacing $\partial/\partial \st$ with  the derivatives of the
Gaussian distribution and making a  
change of variable $\spt-\st \rightarrow \spt$ gives us 
\begin{eqnarray}
1+\xi_{s}\ss &=&  \int   d \spt \,  G(\spt,\sigm\spp) \times  
\label{eq:a20}
\\
&\times&\left[\xi_{r}\spp + \left(1  - \frac{\spt \vp\spp} {2 \sig\spp}
\right)^2 - \frac{\vp^2\spp}{4 \sig\spp} \right] 
 \, \,. \nonumber 
\end{eqnarray}

This expresses  the  galaxy two-point correlation function in redshift
space  as a function of there  quantities in real space, namely the
galaxy two-point correlation function $\xi_r$, the  
line-of-sight component of the pair velocity $\vp$ and the dispersion of    
line-of-sight component of the pair velocity $\sig$. Equation
(\ref{eq:a20}) is non-perturbative and it incorporates all the effects
of redshift space distortions. This includes multi-streaming in
redshift space and all other non-linear effects which may arise due to
the mapping from real to redshift space.  

\section{Comparison with linear results.}
 Kaiser (1987) and Hamilton (1992) have analyzed the effect of
linear redshift distortions on the two-point correlation
function. We use $\xi^L_s$ to refer to this 
and in this section we compare $\xi^L_s$  with $\xi_s$ which incorporates
non-linear redshift distortions.

The expression for $\xi^L_s$ is obtained by keeping only the linear
terms in equation (\ref{eq:a14}) which gives 
\begin{equation}
\xi^L_s\ss=\xi_r\ss - \frac{\partial}{\partial \st} \vp\ss 
+ \frac{1}{2} \frac{\partial^{2}}{\partial \st^{2}} 
\sig\ss 
\label{eq:b1}
\end{equation}
In the linear regime, under the assumption of linear bias,  the real
space quantities $\xi_r$, $\vp$ and  $\sig$  can be expressed in terms
of  $\xi$, the two-point correlation function of the underlying
dark matter distribution (details in Appendix A), as  
\begin{equation}
\xi_r\ss=b^2 \xi(s)
\label{eq:b2}
\end{equation}
where $b$ is the linear bias parameter and $s=\sqrt{s^2_{\perp}+\st^2}$.
We also have  
\begin{equation}
{\vp\ss}=-2 \, \fo \,  b \, \frac{\partial}{\partial \st}
(\nabla^2)^{-1} \xi(s) = 
-\frac{2}{3} \st \, \fo  \, b \,  \xib_2(s) 
\label{eq:b3}
\end{equation}
and
\begin{eqnarray}
\sig\ss&=& 2 \left[ \fosq \frac{\partial^{2}}{\partial \st^{2}}
(\nabla^2)^{-2} \xi(s)  +  \sigma^2 \right] \nonumber \\ 
&=& \fosq \left[ \frac{s^2}{3} \xib_1(s) - \frac{s^2_{\perp}}{3}
\xib_2(s) + \frac{(s^2 - 3 \st^2)}{15} \xib_4(s) \right] 
\label{eq:b4}
\end{eqnarray}
where $\fo \approx \Omega^{0.6}$ is the dimensionless growth
rate for linear perturbation (Peebles 1980), $\sigma^2$ is the
one-dimensional peculiar velocity dispersion  and 
\begin{equation}
\xib_n(s)=\frac{n+1}{s^{n+1}} \int_0^s \xi(y) y^n dy\,.
\label{eq:b5}
\end{equation}
  We use equations (\ref{eq:b2}),(\ref{eq:b3}) and (\ref{eq:b4})
to express both $\xi^L_s$ (eq. \ref{eq:b1}) and $\xi_s$
(eq. \ref{eq:a20}) 
in terms of only three inputs namely $\fo$,$b$ and $\xi$.
This gives an  operator equation (eq. 4 of Hamilton 1992)  
\begin{equation}
\xi^L_s\ss=\left[ b + \fo  \frac{\partial^{2}}{\partial
\st^{2}} (\nabla^2)^{-1} \right]^2  \xi (s)
\end{equation}
for the linear redshift two-point correlation $\xi^L_s$ in terms of
$\xi$ the actual two-point correlation of the underlying dark matter
distribution. 

An alternative way  to parametrize the anisotropy of the redshift
two-point correlation function is to use $s$  and $\mu=\st/s$ (the
cosine 
of the angle between $\s$ and the line of sight $\n$)  instead of
$\ss$. The angular dependence of  $\xi^L_s(s,\mu)$ can be  very
conveniently expressed as a sum of spherical harmonics (Hamilton 1992)
as   
\begin{equation}
\xi^L_s(s)=\xi^L_0(s) P_0(\mu) + \xi^L_2(s) P_2(\mu)
+ \xi^L_4(s) P_4(\mu)
\label{eq:b6}
\end{equation}
where $P_l(\mu)$ are the Legendre Polynomials and $\xi^L_0$, $\xi^L_2$
ans $\xi^L_4$ are the monopole, quadrupole and the hexadecapole
components of the linear redshift two-point correlation function. In
the linear analysis 
these 
\begin{equation}
\xi^L_0(s)=b^2(1+\frac{2}{3} \beta  + \frac{1}{5} \beta^2) \xi(s)
\label{eq:b7}
\end{equation}
\begin{equation}
\xi^L_2(s)=b^2 (\frac{4}{3} \beta + \frac{4}{7} \beta^2) (\xi(s)-\xib_2(s))
\label{eq:b8}
\end{equation}
\begin{equation}
\xi^L_4(s)=\frac{8}{35} \beta^2 b^2 \, [\xi(s) + \frac{5}{2} \xib_2(s)
-\frac{7}{2} \xib_4(s) ];
\label{eq:b9}
\end{equation}
are the only non-zero harmonics, and here $\beta=f/b$. The two-point
correlation function of the underlying dark matter distribution is
largely undetermined and equation (\ref{eq:b7}) cannot be used to
determine both $b$ and $\beta$. 
The ratios of $\xi^L_0$,$\xi^L_2$ and $\xi^L_4$ can be used to only 
determine $\beta$. It is a limitation of the linear analysis that 
$b$ and $\fo$ cannot be determined individually.   

The angular dependence of $\xi_s(s,\mu)$ can also be expanded in terms
of spherical harmonics 
\begin{equation}
\xi(s,\mu)=\sum_{l=0}^{l=\infty} \xi_l(s) P_l(\mu) \,.
\label{eq:b11}
\end{equation}
and here,  as in the linear analysis, the odd terms are all
zero. However in this
case there are no simple analytic expressions for the spherical
harmonics and these have to numerically evaluated. 

We have used a COBE normalised (Bunn and  White, 1996) CDM power
spectrum (Efstathiou, Bond and White, 1992) with shape parameter  
$\Gamma=0.25$ and $h=0.5$ to calculate the real space two-point
correlation function for the underlying dark matter distribution. This
model predicts $\sigma_8=0.55$ and we  expect non-linear effects to
be small in real space at scales $\ge 10 \, h^{-1} {\rm Mpc}$.  Below
we compare the spherical harmonics of  $\xi_s(s,\mu)$ and
$\xi^L_s(s,\mu)$  for this model  for different values of $f$
and $b$.  

\begin{figure}
\includegraphics[angle=-90, width=0.7 \textwidth]{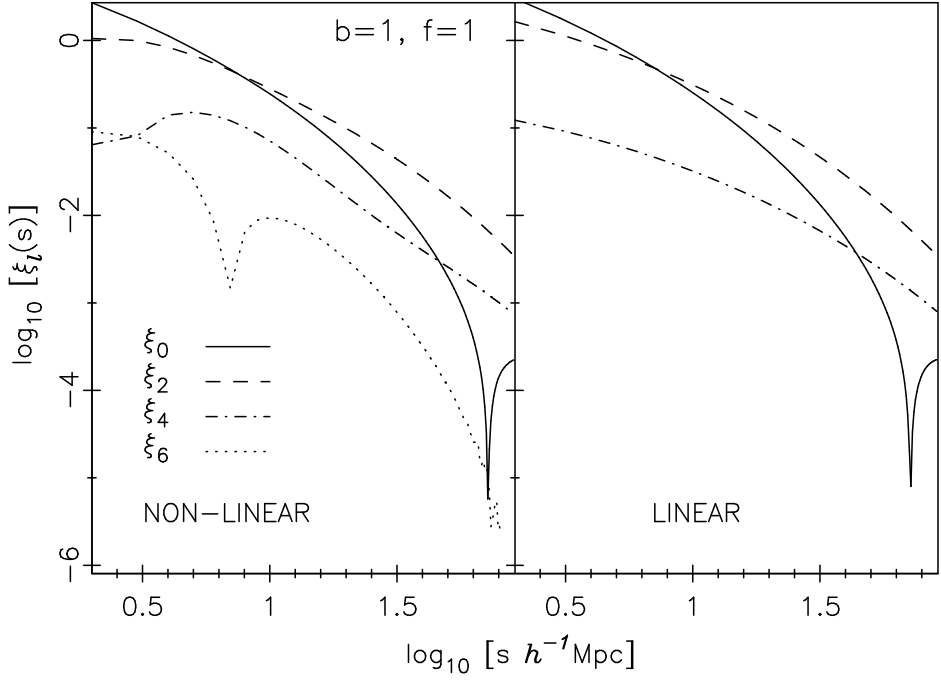} 
\caption{This shows $\xi_l$, the various spherical harmonics of
the redshift space two-point correlation function calculated using
eq. (\ref{eq:a20}) (NON-LINEAR) and  eqs. (\ref{eq:b7})-(\ref{eq:b9})
(LINEAR) for a COBE normalised CDM power spectrum  with $\Gamma=0.25$
and $h=0.5$.} 
\label{fig:1}
\end{figure}

We first consider the case   $f=1,\, b=1$  for which the various
spherical harmonics are shown in figure \ref{fig:1}. We find that
the deviations from the linear predictions are more pronounced for
the higher spherical harmonics. While the behaviour of the monopole
$\xi^L_0$  shows practically no deviations from $\xi^L_0$,  $\xi_4$
exhibits significant deviations from s $\xi^L_4$.  
Another feature is that the non-linear redshift
distortions produce  a non-zero $\xi_6$ whereas the linear analysis
predicts a value of $zero$ for this and all higher harmonics. We have
not considered any of the higher harmonics here.      

\begin{figure}
\includegraphics[angle=-90, width=0.7 \textwidth]{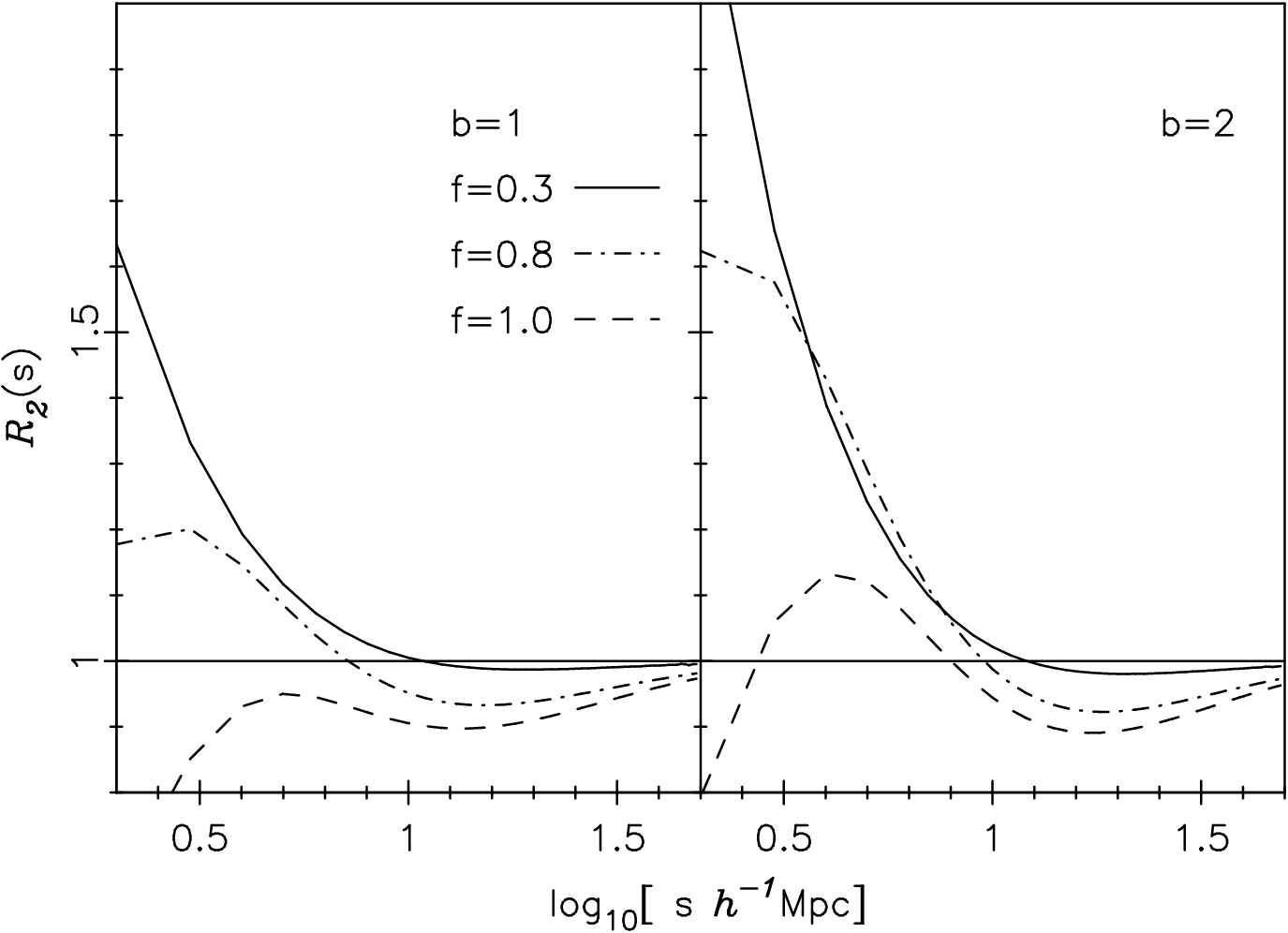} 
\caption{The plots show $R_2=\xi_2/\xi^L_2$ for different values of
$\fo$ and $b$ for a COBE normalised $\Gamma=0.25$, $h=0.5$ CDM power
spectrum.}
\label{fig:2}
\end{figure}

\begin{figure}
\includegraphics[angle=-90, width=0.7 \textwidth]{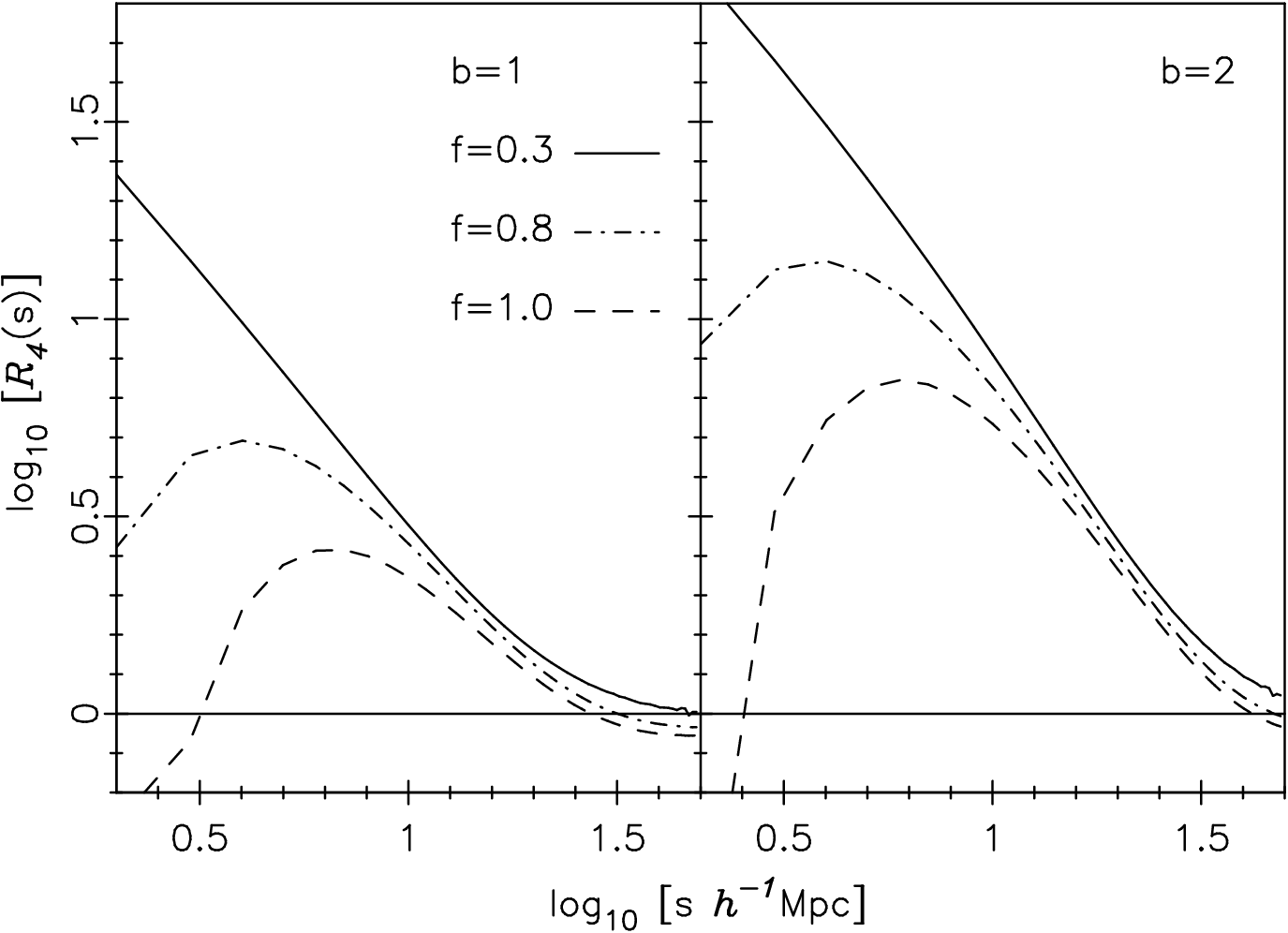} 
\caption{The plots show $R_4=\xi_4/\xi^L_4$ for different
values of $\fo$ and $b$ for a COBE normalised $\Gamma=0.25$, $h=0.5$
CDM power spectrum.}  
\label{fig:3}
\end{figure}

We next analyze the behaviour of $\xi_2$ and $\xi_4$ in some more
detail using the quantity   $R_l(s)=\xi_l(s)/\xi^L_l(s)$  whose 
deviation from the value  $R_l(s)=1$ indicates departures from the
linear predictions.  The behaviour of $R_2$ and $R_4$ are shown in
figures \ref{fig:2} and \ref{fig:3}. We see that the behaviour of $R_2$
and $R_4$ looks  similar, the difference being that the magnitude
of the deviations from the linear predictions is much larger for the
latter. A feature common to $R_2$ and $R_4$  is that the deviations
from the linear predictions increase monotonically as we go to smaller
length-scales for low $\fo$, , whereas for larger 
$\fo$ the deviations saturate  and even fall at smaller
scales. A possible explanation is that  at small scales the
anisotropies in $\xi_s$ are erased by multi-streaming in redshift
space. The peculiar velocities increase with $\fo$ and we can expect
the effects of multi-streaming to also increase with $\fo$. 
 We find that $\xi_4$ is around
a factor of two larger than $\xi^L_4$ on scales $10 \, - 20 \, h^{-1}
{\rm Mpc}$ where the non-linear effects are expected to be small
in real space. The value of $R_4$ increases at scales smaller than $10
h^{-1} {\rm Mpc}$ but  real space non-linear effects are also
expected to become significant at these scales and our results are not
expected to give the true picture here. Both $R_2$ and $R_4$  
approach one at large scales.  

\begin{figure}
\includegraphics[angle=-90, width=0.7 \textwidth]{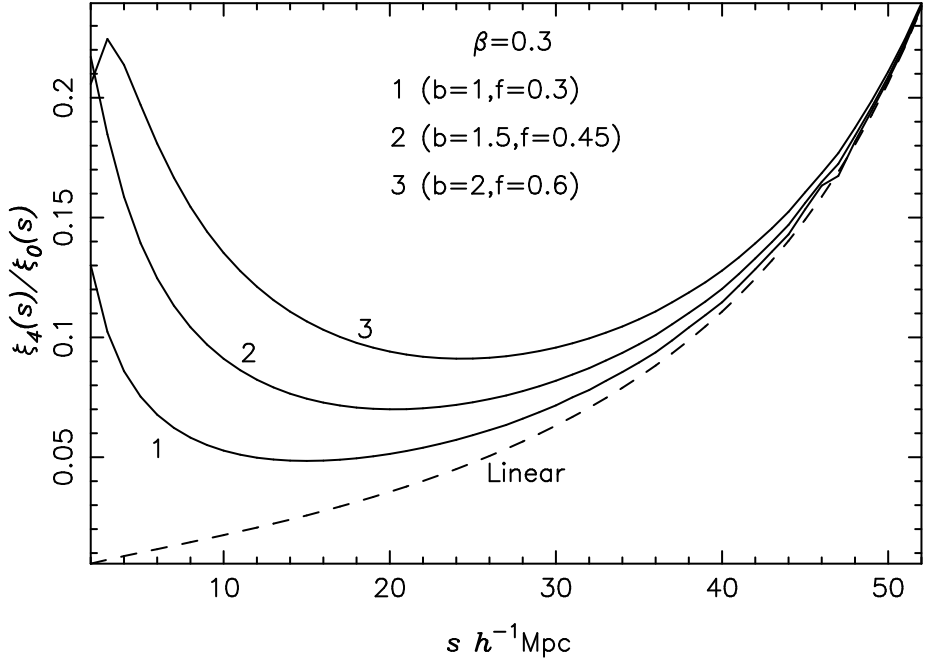} 
\caption{This shows $\xi_4/\xi_0$ for different
values of $\fo$ and $b$ all with $\beta=\fo/b=0.3$ for a COBE normalised
$\Gamma=0.25$, $h=0.5$ CDM power spectrum.}  
\label{fig:4}
\end{figure}

The  $b$ and $\fo$ dependence  of the non-linear redshift space
distortions cannot be expressed through just one parameter
$\beta$. We consider the behaviour of the ratio
$\xi_4(s)/\xi_0(s)$ (figure \ref{fig:4}) to study this effect. Linear
redshift distortions (eqs. \ref{eq:b7} and \ref{eq:b9}) predict this  
to depend on just $\beta$.   We see that the our predictions for
$\xi_4(s)/\xi_0(s)$ 
change considerably if we vary $\fo$ and $b$ keeping $\beta$
fixed. This holds the possibility of using the redshift two-point 
correlation function to determine $\fo$ and $b$ separately. 

\section{Discussion and summary.}
We have calculated
 $\xi_s$ the galaxy two-point correlation function in 
redshift space for a situation where the density fluctuations and
peculiar velocities  are linear in real space. Our calculation
incorporates all non-linear effects which arise due to  the mapping
from real to redshift space.  Our result (equation \ref{eq:a20}) is
non-perturbative and it  includes the effects  of multi-streaming in
redshift space.  

The position of
a galaxy in redshift space is a combination of its actual position and
the line of sight  component of its peculiar velocity, and the
redshift two-point correlation function will have contributions from
three effects 
\begin{itemize}
\item[1.] Correlations between the actual positions of the
galaxies. This is  quantified by the real space two-point correlation
$\xi_r(s)$.  
\item[2.] Correlations between the peculiar velocities and the
actual positions of galaxies. This is quantified by the mean pair
velocity whose component along the light of sight  is $\vp(\s)$. In
the 
linear  regime this is negative (eq \ref{eq:ap3}) because of coherent 
flows out of  under dense regions and into over dense regions.  
\item[3.] Correlations between the peculiar velocities of galaxies. 
This is quantified by the pair velocity dispersion 
whose component along the light of sight  is 
$\sig(\s)=2 [\sigma^2 - \langle \n \cdot \v(\s_1) \n \cdot
\v(\s_2) \rangle]$  where $\s=\s_2-\s_1$. Here $\sigma^2$ is the one
dimensional  peculiar velocity dispersion of the random motion of
individual galaxies and $\langle \n \cdot \v(\s_1) \n \cdot
\v(\s_2) \rangle = - f^2 \frac{\partial^2}{\partial s^2_{\parallel}}
(\nabla^2)^{-2}  \xi(s)$ (eq. \ref{eq:ap5})  is the correlation in the
line of sight 
component of peculiar velocities due to coherent flows. At large
separations the effect of the coherent flows is much smaller than the
contribution from  random motions  and $\sig(\s) \simeq 2 \sigma^2$. 
The velocity-velocity correlations due to coherent flows 
increases at smaller separations. This causes  
$\sig(\s)$ to be anisotropic and its value is less than $2 \sigma^2$. 
In the linear analysis $\sig(\s)$ decreases
monotonically as $s$ is reduced and $\sig(0)=0$. 
\end{itemize}

The linear redshift two-point correlation (eq. \ref{eq:b1})
incorporates these three effects and   $\xi^L_s(\s)$ at a separation
$\s$ in redshift space is  expressed in terms of $\xi_r(\s)$,$\vp(\s)$
and  $\sig(\s)$  at the 
same separation in real space.  The line of sight component of the
relative peculiar velocity $U$ between a pair of galaxies causes their 
separation in redshift space to be 
different from the actual separation, and in principle the real space
quantities should be evaluated at a different separation $\r$ where
$r_{\parallel}=\st -U$ and $r_{\perp}=s_{\perp}$. A possible way to
incorporate this effect is to use (eq.  \ref{eq:b1}) for $\xi_s^L(\s)$
and evaluate all the real space quantities at $\r$, assuming that  the 
different values of $U$ 
being given by a  probability  distribution $f(U)$. This gives 
\begin{equation}
\xi_s\ss = \int d U f(U) \xi_s^L(s_{\perp},s_{\parallel}-U) \,.
\label{eq:dis1}
\end{equation}
The distribution of $U$ is characterized by $\sig\ss$ which in general
is anisotropic and  scale dependent. As a simplifying assumption 
the distribution of $U$ is usually taken to be isotropic and the 
function $f(U)$  is  chosen to be either a Gaussian or an
exponential with only one constant parameter  $2 \sigma^2$ which is
the value of $\sig(\s)$ at  large separations.  
This phenomenological model for the non-linear redshift two-point
correlation has been found to match the results of N-body
simulations. Two effects which are not included in this
model are  (1.) the anisotropy and spatial dependence of $\sig(\s)$
arising from the coherent flows, and (2.)  the fact that the
distribution of $U$ is  correlated  with the density fluctuations.  

Our exact calculation takes into account all of the effects discussed
above. We find that the redshift two-point correlation function
(eq. \ref{eq:a20})  can be expressed as  the real space two-point
correlation function combined with ratios of the pair velocity to the
pair  velocity dispersion, convolved along  the line of sight with a
Gaussian one dimensional pair  velocity distribution. To get a better
understanding of the various terms in (eq. \ref{eq:a20}) we first
consider a limiting situation where the separation $s$ is taken to be
large so that $\langle \n \cdot \v(\s_1) \n \cdot
\v(\s_2) \rangle \ll \sigma^2$ i.e.   
the contribution to $\sig(\s)$ from the random
motions is much larger than  the contribution from  coherent
flows.  In this situation  we can ignore the spatial variation of
$\sig(\s)$ (except for  the contribution of the derivative of $\sig(\s)$
to $\xi^L_s(\s)$). The calculation of $\xi_s$ in this limit is presented in
Appendix B. We find that in this limit the result of our
calculation (eq. \ref{eq:apb2}) is identical to the phenomenological
model (eq. \ref{eq:dis1}) with a Gaussian pair velocity distribution. 
This can be interpreted in terms of a simple
picture  where  all the non-linear effects of the redshift distortions
on $\xi_s$  arise from the random motions of galaxies. The random
rearrangement of the galaxy positions over the length scale $\sigma$
along the line of sight in redshift surveys modulates the linear
redshift correlation function $\xi^L_s(s)$. This  
diffusion process acting on $\xi^L_s(\s)$ along the line of sight  
introduces an elongation  in $\xi_s$. 

It is interesting to note that a similar interpretation is also
possible  for the non-linear effects on  the two-point correlation
function in the Zel'dovich approximation (Bharadwaj 1996b) and in the
perturbative treatment of gravitational instability (Bharadwaj
1996a). In both situations the scaling properties of the non-linear
effects is determined by the pair velocity dispersion (Bharadwaj 1997) 
and at large scales the non-linear two-point correlation function can
be interpreted in terms of the linear two-point  correlation function
modified by  small scale random motions. 

We now come back to the interpretation of equation
(\ref{eq:a20}). One of the main features is that it incorporates all
the effects of the  anisotropy and scale dependence of the   pair
velocity dispersion.  As noted earlier, these two features arise
because of the coherent flows.  Whereas the convolution in equation
(\ref{eq:dis1}) accounts only for  the effects of random motions,
the convolution with the distribution function in equation
(\ref{eq:a20}) also includes the effect of the coherent flows. The 
coherent  flows reduce the width of the distribution function, causing a
reduction in  the elongation in $\xi_s$. The effect of the coherent
flows is also present in the terms involving  ratios of $\vp$ and
$\sig$, the contribution of these terms increasing with the effect of 
coherent flows.  All the terms involving $\vp^2$ arise due to the
non-linear redshift distortions.    

In the linear analysis the redshift distortions produced by the
coherent flows give rise to a flattening of the two-point correlation
function along the line of sight. This effect is best captured by the
quadrupole moment $\xi_2$ which is predicted to be negative. The
linear analysis also predicts a positive hexadecapole moment arising
from the term involving the second derivative of $\sig(\s)$ in
equation (\ref{eq:b1}) for $\xi^L_s(\s)$. 

Comparing our result with the predictions of the linear analysis of
redshift distortions we see that the non-linear redshift space effects
 are more pronounced on the higher spherical moments $\xi_l$ of
the redshift space two-point correlation function. While the monopole
$\xi_0$ shows no significant deviations from the linear predictions,
the deviations are around $10 \%$ on scales of $10 \,- \,  20 \,
h^{-1} {\rm Mpc}$ for $\xi_2$. The moment $\xi_4$ shows very large
deviations from 
the linear predictions, the non-linear predictions differing  by a
factor of $\sim 2$ on scales  of $10 \,- \,  20 \, h^{-1} {\rm Mpc}$.   

A point which should be noted is that our analysis ignores all 
non-linear effect in the real space density fluctuations and peculiar
velocities, and focuses only on the non-linear redshift 
distortions. While this assumption is expected to be valid for $\xi_r$
and $\vp$ at scales greater than $10 h^{-1} {\rm Mpc}$, we do not expect
this to hold for $\sig$ which has a constant isotropic part $2
\sigma^2$ arising from the random motions. This has a large
contribution 
from small scales which are non-linear and we do not expect the 
linear results used here to give a realistic estimate of $\sig$.
We also expect this effect to be present in our estimates on the
effects on non-linear redshift distortions on the different angular
moments of $\xi_s$.   A possible solution is to use N-body simulations
to determine the  constant, isotropic part of $\sig$ and to use linear
theory to  calculate the contribution of the coherent flows.  
This possibility has not been explored here, and we plan to
continue future work in this direction.

 Another possibility which
might arise but has not been taken into account is  non-linear
bias. A quadratic bias relation gives rise to two bias parameters
$b_1$ and $b_2$, and the relation between the fluctuations in the
galaxy number density and the matter density is 
$\delta_g=b_1 \delta_m + \frac{1}{2} b_2 \delta_m^2$. In this
situation $\delta_g$ will no longer be a Gaussian random field and
this will give rise to a new series of terms in equation
(\ref{eq:a14}) which could be summed up following the same procedure.  

The effects of linear redshift distortion can be  parametrized by a
single parameter $\beta=\fo/b$ and the analysis of redshift surveys
based on this do not yield $\fo$ and $b$ separately. We find that
$\xi_4$, the  hexadecapole  moment of $\xi_s$, depends  on
$\fo$ and $b$ separately when the non-linear effects of redshift
distortions are taken into account. This holds the possibility of
separately determining  $\fo$ and $b$ from redshift surveys. Although
this effect increases at small scales, the
range $10-20 h^{-1} {\rm Mpc}$ is possibly best suited for putting  
it to use. Non-linear effects in real
space and non-linear biasing will become important at smaller scales. 
At these scales the hexadecapole moment is expected to be 
$10 \%$ of the quadrupole moment. Further investigations are needed 
to determine the range of validity of the results presented here
before they can be applied to the analysis of redshift surveys. 

\section{Acknowledgements.}
The author would like to thank Debaprasad Giri and Akhilesh Singh for
their help while writing this paper. The author is also thankful to
MRI, Allahabad for its hospitality during a part of the period when
this was being completed. 
\appendix
\section{}
In this appendix we discuss, in some detail, the relation between
$\xi_r(r)$, $\vp(\r)$, $\sig(\r)$ and $\xi(r)$.   In the linear regime 
$\delta(\ro)$, the fluctuation in the dark matter
density, can be expressed in terms of a potential $\psi(\ro)$ as
\begin{equation}
\delta(\ro)=\nabla^2 \psi(\ro) \, , 
\end{equation}
and the peculiar velocity $\v(\ro)$ can be expressed as 
\begin{equation}
\v(\ro)=-\fo \grad   \psi(\ro) \, ,  
\end{equation}

In addition, assuming a linear bias relation,  $\delta_g(\ro)$ the
fluctuation in the galaxy number density can be expressed as 
\begin{equation}
\delta_g(\ro)=b \nabla^2 \psi(\ro) \,.
\end{equation}

Using these we can express $\xi(r)$ as 
\begin{equation}
\xi(r)=\langle \delta(\ro) \delta(\rw) \rangle = 
\langle \nabla^2 \psi(\ro) \nabla^2\psi(\rw) \rangle  = \nabla^4 \phi(r)
\label{eq:ap1}
\end{equation} 
where $\r =\rw-\ro$ and $\phi(r)= \langle \psi(\ro) \psi(\rw) \rangle$. 
Equation (\ref{eq:ap1}) can be inverted  to express $\phi(r)$ and its
derivatives in terms of $\xi(r)$. 

We use these to calculate $\vp(\r)$,  which gives us 
\begin{eqnarray}
\vp(\r) &=& \langle \n \cdot (\v(\rw)-\v(\ro))
(\delta_g(\rw)+\delta(\ro))  \rangle \nonumber \\
&=& (-b f)\langle \n \cdot ( \grad \psi(\rw) - \grad \psi(\ro)) ( \nabla^2
\psi(\rw)+  \nabla^2 \psi(\ro)) \rangle \nonumber \\
&=& -2f b (\n \cdot \grad) \nabla^2 \phi(r)
\label{eq:ap2}
\end{eqnarray}
This can be further simplified using  equation (\ref{eq:ap1}) to
 obtain 
\begin{eqnarray}
\vp(\r)&=& -2 f b \frac{\partial}{\partial r_{\parallel}}
(\nabla^2)^{-1} \xi(r) \nonumber \\
&=& -\frac{2 b f r_{\parallel}}{3}  \xib_2(r)
\label{eq:ap3}
\end{eqnarray}

We next consider $\sig(\r)$ which can be expressed as 
\begin{eqnarray}
\sig(\r)  &=&  \langle \n \cdot (\v(\rw)-\v(\ro)) \, \,  \n \cdot
(\v(\rw)-\v(\ro))  \rangle \nonumber \\
&=& - 2 f^2 <[\n \cdot \grad \psi(\ro) ] [\n \cdot \grad \psi(\rw) ]
\rangle + \sigma^2  \nonumber \\
&=& 2 f^2  \frac{\partial^2 }{\partial r^2_{\parallel}} \phi(r) +
\sigma^2 
\label{eq:ap4}
\end{eqnarray}
where $\sigma^2=<(\n \cdot \v)^2>$ is the one dimensional peculiar
velocity dispersion.  The expression for $\sig(\r)$ can be further
simplified using  (\ref{eq:ap1}) to obtain 
\begin{eqnarray}
\sig(\r)=&=& 2 f^2  \frac{\partial^2 }{\partial r^2_{\parallel}}
(\nabla^2)^{-2} \xi(r)+\sigma^2 \nonumber \\
&=& f^2 \left[ \frac{r^2}{3} \xib_1(r) - \frac{r^2_{\perp}}{3}
\xib_2(r) + \frac{(r^2 - 3 r^2_{\parallel})}{15} \xib_4(r) \right]
\label{eq:ap5}
\end{eqnarray} 

\section{}
In this appendix we consider a limiting situation where the separation
$s$ at which the redshift two-point correlation is being calculated is
very large. We also assume that the real space two-point correlation
function  is a power law $\xi_r(s) \propto x^{-\gamma}$ with
$\gamma>2$ at large separations. The power law requirement is not
crucial and the arguments presented here are valid provided the
integral $\int^{x}_{y} \xi_r(s) s d s$ converges for $x \rightarrow
\infty$.   

With the power law assumption we have $\vp(\s) \propto s^{1-\gamma}$
(eq. \ref{eq:ap3}), $\sig(\s) - 2 \sigma^2 \propto s^{2-\gamma}$
(eq. \ref{eq:ap5}) and $\sig(\s) \simeq 2 \sigma^2$ for large
values of $s$. The point to note is that at large separations the 
correlation between the galaxy peculiar velocities is very small and
the pair velocity dispersion is nearly isotropic and has a constant
value $2 \sigma^2$. 

We next shift our attention to equation (\ref{eq:a14})  which expresses
$\xi_s$ 
as a series of different terms involving $\xi_r$,$\vp$ and $\sig$. In 
the linear regime each  of these quantities  is 
characterized by a small number $\epsilon \sim \delta \rho/\rho$ and 
$\xi_r \sim \vp \sim \sig \sim \epsilon^2$. Identifying all the terms
of order $\epsilon^{2m}$ in  equation (\ref{eq:a14}) we have 
\begin{eqnarray}
& &\frac{1}{ 2^m m!} \frac{\partial^{2m}}{\partial \st^{2m} \, }
[\sig(\s)]^{m} 
+ \frac{1}{ 2^{m-1} (m-1)!} \frac{\partial^{2m-2}}{\partial
\st^{2m-2}} \,  [\sig(\s)]^{m-1} \xi_r (\s) \nonumber \\ 
&+&  \frac{1}{ 2^{m-1} (m-1)!} \frac{\partial^{2m-1}}{\partial
\st^{2m-1}} \,  [\sig(\s)]^{m-1} \vp (\s) \nonumber \\ 
&+&  \frac{1}{ 2^{m-2} (m-2)!} \frac{\partial^{2m-2}}{\partial
\st^{2m-2}} \,  [\sig(\s)]^{m-2} \frac{\vp^2(\s)}{4} \sim \epsilon^{2m}
\end{eqnarray}
This contains  terms involving different powers of $s$. The terms
whose $s$ dependence is $s^{2-\gamma-2m}$ fall off slowest with
increasing $s$ and these terms 
dominate at large separations. Retaining only these terms the above
expression becomes 
\begin{eqnarray}
\frac{[\sigma^2]^{m-1}}{(m-1)!}  \left[ \frac{1}{2}
\frac{\partial^{2m}}{\partial 
\st^{2m} }\, \sig(\s) +
 \frac{\partial^{2m-2}}{\partial \st^{2m-2}} \, \xi_r(\s) 
-  \frac{\partial^{2m-1}}{\partial \st^{2m-1}} \, \vp(\s) \right]
\end{eqnarray} 
Summing up all the terms with different powers of $\epsilon^2$ we 
obtain  
\begin{eqnarray}
1+\xi_s(\s)&=1+ &\sum_{q=0}^{\infty}
\frac{\sigma^{2q}}{q!} \frac{\partial^{2q}}{\partial \st^{2q}} \left[ 
 \xi_r(\s) -   \frac{\partial}{\partial \st}  \vp(\s) 
+  \frac{1}{2} \frac{\partial^2}{\partial \st^2} \sig(\s)
 \right] \label{eq:apb1}   
\end{eqnarray}
where the terms in the square bracket is the linear redshift two-point
correlation function $\xi^L_s(\s)$ (equation \ref{eq:b1}). We next sum
up all the terms following  a procedure similar to that used in section 2
(equation \ref{eq:a15} to \ref{eq:a20}) and  obtain 
\begin{eqnarray}
\xi_s\ss=\int d \st^{'} \,  G(\st^{'},\sqrt{2} \sigma) \,  \xi^L_s
\spp \,. 
\label{eq:apb2}
\end{eqnarray}

\begin{thebibliography}{11}
\bibitem{bal} Ballinger W.E., Peacock J. A., Heavens A. F., 1996,
\mnras, 282, 877 
\bibitem{ben} Benson A. J., Cole S., Frenk C. S., Baugh C. M.,
Lacey C. G., 2000, \apj, 311, 793
\bibitem{bh4} Bharadwaj S., 1996a, \apj, 460, 28
\bibitem{bh3} Bharadwaj S., 1996b, \apj, 472, 1 
\bibitem{bh2} Bharadwaj S., 1997, \apj, 477, 1 
\bibitem{bh1} Bharadwaj S.,  1999, \apj, 516, 507 
\bibitem{Brainerd1994} Brainerd T. G., Bromley B. C., Warren M. S.,
Zurek W. H., 1996, \apj, 464, L103 
\bibitem{bromley1997} Bromley B. C., Warren M. S., Zurek W. H., 1997,
\apj, 475, 414  
\bibitem{6} Bunn E. F.,  White M. 1996, \apj, 460, 107
\bibitem{cen} Cen R., Ostriker J. P., 1992. \apj, 399, L113 
\bibitem{Dek} Dekel A., Lahav O., 1999, \apj, 520, 24
\bibitem{d5}  Dressler A., 1980, \apj, 236, 351
\bibitem{26}  Efstathiou G., Bond J. R.,  White S. D. M., 1992,
\mnras, 250, 1p
\bibitem{x6} Fisher K. B., Davis M., Strauss M. A., Yahil A., Huchra
J. P., 1994, \mnras, 267, 927
\bibitem{x7} Fisher K. B.,Scharf C. A., Lahav O., 1994, \mnras, 266, 219 
\bibitem{Fisher1006} Fisher K.B., Nusser A. 1996, MNRAS, 279L, 1
\bibitem{x10} Gramman M., Cen R., Bahcall N. A., 1993, \apj, 419, 440
\bibitem{guz} Guzzo L., Strauss M. A., Fisher K. B., Giovanelli R., 
Haynes, M. P., 1997, \apj, 489, 37
\bibitem{x11} Hamilton A J. S., 1992, \apj, 385, L5  
\bibitem{Hamilton1998} Hamilton A. J. S., 1998, in The Evolving Universe,
ed. D. Hamilton, Kluwer,Dordrecht, p185
\bibitem{x12} Hamilton A. J. S., Culhane M., 1996, \mnras, 278, 73
\bibitem{Hatton1998} Hatton S. J., Cole S., 1998, MNRAS, 296, 10
\bibitem{x13} Heavens A. F.,  Taylor A. N., 1995, \mnras, 275, 483
\bibitem{} Hermit S., Santiago B. X., Lahav O., Strauss M. A.,
Davis M., Dressler A.,  Huchra, J. P., 1996, \mnras, 283, 709
\bibitem{Hivon1995} Hivon E., Bouchet F.R., Colombi S., 
Juszkiewicz R., 1995, A\&A, 298, 643  
\bibitem{Hui2000} Hui L., Kofman L., Shandarin S.F., 2000, 537, 12
\bibitem{x14} Kaiser N., 1987, \mnras, 227, 1
\bibitem{kauf} Kauffmann G. Nusser A., Steinmetz, M., 1997,
\mnras, 286, 795
\bibitem{Lah} Lahav O., Nemiroff R. J., Piran T., 1990, \apj, 350, 119
\bibitem{Lov} Loveday J., Maddox S. J., Efstathiou G., Peterson, B. A., 
1995, \apj , 442, 457
\bibitem{mann} Mann R. G., Peacock J. A., Heavens A. F., 1998,
\mnras, 293, 209
\bibitem{melott} Melott A. L., Coles P., Feldman, H. A., Wilhite B.,
 1998, \apj, 496, L85    
\bibitem{Peacock2} Peacock J. A., Dodds, S.J., 1994, \mnras, 267, 1020
\bibitem{Peacock1} Peacock, J. A. \etal,  2001, Nature, 410, 169
\bibitem{x21} Peebles P. J. E., 1990, The LArge-Scale Structure of
the Universe, Princeton University Press, Princeton, NJ
\bibitem{pen} Pen U.-L., 1998, \apj, 504, 601
\bibitem{San} Santiago B. X., Strauss, M. A., 1992, \apj, 387, 9
\bibitem{strauss} Strauss M. A., Willick J. A., 1995, Phys. Rep,
261, 271
\bibitem{1991ApJ...370L..15S} Suto Y.,  Suginohara T.,
1991,\apjl, 370, L15    
\bibitem{Szalay} Szalay A.S., Matsubara T., Landy, S.D.,  1998, 
ApJ, 498, L1
\bibitem{taruya} Taruya A., Magara H., Jing, Y. P.,  Suto Y., 2001,
PASJ, 53, 155   
\bibitem{Taylor 1996} Taylor A. N., Hamilton, A. J. S.,  1996, MNRAS,
282, 767 
\bibitem{wu} Wu J. H. P., 2000, preprint, astro-ph/0012207 
\bibitem{x24} Zaroubi S., Hoffman Y., 1996, 462, 25
\end{thebibliography}
\end{document}